\documentclass[sigconf,natbib=true]{acmart}

\usepackage{bigstrut}
\usepackage{booktabs}
\usepackage{multirow}
\usepackage{microtype}
\usepackage{graphicx}
\usepackage{subfig}
\usepackage{overpic}
\usepackage{balance}
\usepackage{amsthm}
\usepackage{array}
\usepackage{clrscode}
\usepackage{multicol}
\usepackage{float}
\usepackage{newfloat}
\usepackage{color}
\usepackage{xcolor}
\usepackage{amsopn}
\usepackage{mathrsfs}
\usepackage{mathtools}
\usepackage{amsmath}
\usepackage{arydshln}
\usepackage{blkarray}
\usepackage{enumerate}
\usepackage{courier}
\usepackage{rotating}
\usepackage{bm}
\usepackage{wrapfig}
\usepackage{ragged2e}
\usepackage[misc]{ifsym}
\usepackage{threeparttable}
\usepackage{makecell}
\usepackage{adjustbox} 
\usepackage{enumitem}
\usepackage{hyperref}
\usepackage{url}
\usepackage{xspace}
\usepackage{comment}
\usepackage{changepage}
\usepackage{algorithm}
\usepackage{algpseudocode} 
\usepackage{algorithmicx} 
\usepackage[most]{tcolorbox}
\usepackage[ruled, vlined, nofillcomment, linesnumbered, algo2e]{algorithm2e}

\definecolor{codegreen}{rgb}{0,0.3,0.6}
\definecolor{codegray}{rgb}{0.5,0.5,0.5}

\newcommand{\ie}{\emph{i.e.,}\xspace}
\newcommand{\eg}{\emph{e.g.,}\xspace}

\newcommand{\paratitle}[1]{\vspace{1.5ex}\noindent\textbf{#1}}
\newcommand{\wrt}{w.r.t.\xspace}

\newcommand{\ignore}[1]{}

\newcommand{\method}{DeepRec\xspace}

\AtBeginDocument{%
  \providecommand\BibTeX{{%
    \normalfont B\kern-0.5em{\scshape i\kern-0.25em b}\kern-0.8em\TeX}}}

\setcopyright{acmlicensed}
\copyrightyear{2018}
\acmYear{2018}
\acmDOI{XXXXXXX.XXXXXXX}

\acmConference[Conference acronym 'XX]{Make sure to enter the correct
  conference title from your rights confirmation emai}{June 03--05,
  2018}{Woodstock, NY}
\acmISBN{978-1-4503-XXXX-X/18/06}

\begin{document}

\title[DeepRec: Towards a Deep Dive Into the Item Space with Large Language Model Based Recommendation]{DeepRec: Towards a Deep Dive Into the Item Space with \\Large Language Model Based Recommendation}

\author{Bowen Zheng\textsuperscript{*}}
\orcid{0009-0002-3010-7899}
\affiliation{%
    \institution{Renmin University of China}
    \city{Beijing}
    \country{China}
}
\email{bwzheng0324@ruc.edu.cn}

\author{Xiaolei Wang\textsuperscript{*}}
\orcid{}
\affiliation{%
    \institution{Renmin University of China}
    \city{Beijing}
    \country{China}
}
\email{wxl1999@foxmail.com}

\author{Enze Liu\textsuperscript{*}}
\orcid{0009-0007-8344-4780}
\affiliation{%
    \institution{Renmin University of China}
    \city{Beijing}
    \country{China}
}
\email{enzeliu@ruc.edu.cn}

\author{Xi Wang}
\affiliation{%
    \institution{Beijing Institute of Technology}
    \city{Beijing}
    \country{China}
}
\email{xiwangai@bit.edu.cn}

\author{Lu Hongyu}
\affiliation{%
    \institution{WeChat, Tencent}
    \city{Guangzhou}
    \country{China}
}
\email{luhy94@gmail.com}

\author{Yu Chen}
\affiliation{%
    \institution{WeChat, Tencent}
    \city{Beijing}
    \country{China}
}
\email{nealcui@tencent.com}

\author{Wayne Xin Zhao\textsuperscript{\Letter}
}
\orcid{0000-0002-8333-6196}
\affiliation{
    \institution{Renmin University of China}
    \city{Beijing}
    \country{China}
}
\email{batmanfly@gmail.com}

\author{Ji-Rong Wen}
\orcid{0000-0002-9777-9676}
\affiliation{
    \institution{Renmin University of China}
    \city{Beijing}
    \country{China}
}
\email{jrwen@ruc.edu.cn}

\thanks{* \ Equal contribution.} 
\thanks{\Letter \ Corresponding author.}

\renewcommand{\shortauthors}{Bowen Zheng, et al.}

\begin{abstract}

Recently, large language models (LLMs) have been introduced into recommender systems (RSs), either to enhance traditional recommendation models (TRMs) or serve as recommendation backbones.
However, existing LLM-based RSs often do not fully exploit the complementary advantages of LLMs (\eg world knowledge and reasoning) and TRMs (\eg recommendation-specific knowledge and efficiency) to fully explore the item space.
To address this, we propose \textbf{\method}, a novel LLM-based RS that enables \textit{autonomous} multi-turn interactions between LLMs and TRMs for \textit{deep} exploration of the item space.
In each interaction turn, LLMs reason over user preferences and interact with TRMs to retrieve candidate items.
After multi-turn interactions, LLMs rank the retrieved items to generate the final recommendations.
We adopt reinforcement learning~(RL) based optimization and propose novel designs from three aspects: recommendation model based data rollout, recommendation-oriented hierarchical rewards, and a two-stage RL training strategy. 
For data rollout, we introduce a preference-aware TRM, with which LLMs interact to construct trajectory data.
For rewards, we design a hierarchical reward function that involves both process-level and outcome-level rewards to optimize the interaction process and recommendation performance, respectively.
For RL training, we develop a two-stage training strategy, where the first stage aims to guide LLMs to interact with TRMs and the second stage focuses on performance improvement.
Experiments on public datasets demonstrate that \method significantly outperforms both traditional and LLM-based baselines, offering a new paradigm for deep exploration in recommendation systems.
Our code is available at \textcolor{blue}{\url{https://github.com/RUCAIBox/DeepRec}}.

\end{abstract}



\maketitle

\begin{figure}[t]
\centering
\includegraphics[width=\linewidth]{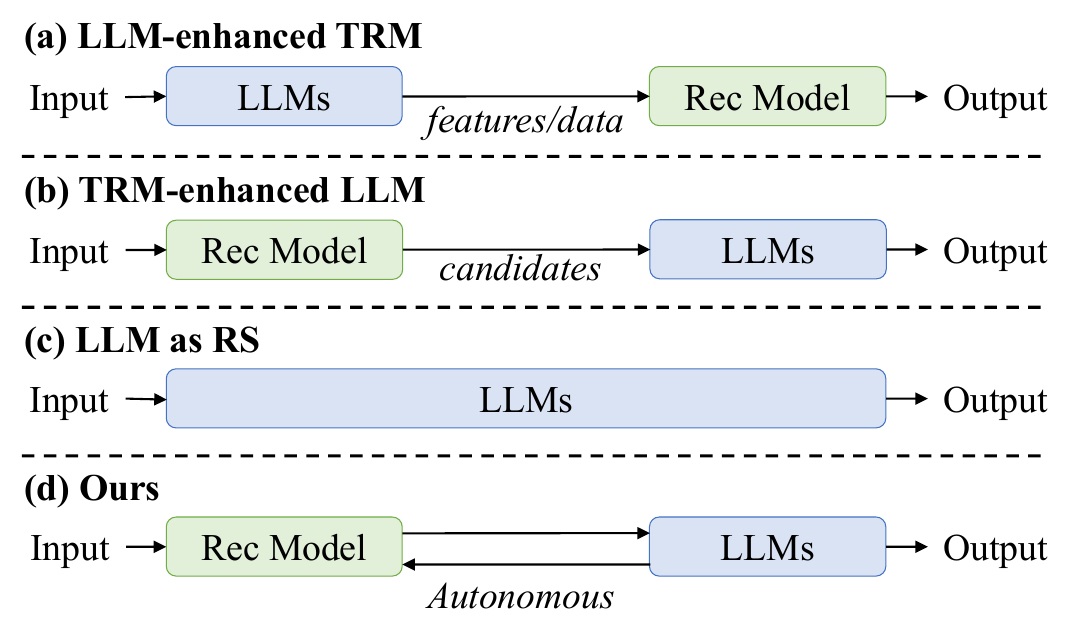}
\caption{
Comparison between existing LLM-based RSs and ours.
}
\label{fig:intro}
\end{figure}

\section{Introduction}
\label{sec:intro}

Nowadays, recommender systems~(RSs) play a crucial role in improving user experience across various online platforms such as e-commerce, news, and micro-video services, owing to their ability to deliver personalized content and resources. 
As user preferences continuously evolve over time, sequential recommendation~\cite{gru4rec,sasrec,bert4rec} has garnered wide research attention, which aims to capture the temporal dynamics of user behaviors and the progression of user interests.
To effectively model such evolving user preferences, sequential recommendation requires not only rich knowledge about each interacted item but also sophisticated reasoning over the entire interaction sequence, which traditional recommendation models~(TRMs) trained on observed data cannot well satisfy.

To solve this, large language models~(LLMs) have been introduced into recommendation~\cite{llmrec_survey1}, which excel at world knowledge and complex reasoning.
Existing LLM-based RSs can be divided into three main categories: \textit{LLM-enhanced TRM}, \textit{TRM-enhanced LLM}, and \textit{LLM as RS}.
{LLM-enhanced TRM}~\cite{kar,unisrec,li2023ctrl} integrates LLMs as auxiliary components to improve TRMs, typically through feature enrichment~\cite{unisrec,kar,llmesr} or data augmentation~\cite{DBLP:journals/corr/abs-2409-13545}.
However, this approach often suffers from misalignment, as general LLMs are not specially optimized for recommendation.
In addition, it cannot fully leverage the reasoning capability of LLMs for user preference modeling.
{TRM-enhanced LLM}~\cite{llamarec,llmrank} extends LLMs with TRMs to generate recommendations.
This approach usually first employs TRMs to retrieve candidate items and then prompts LLMs with these candidates for further ranking.
Although LLMs are aligned with recommendation through TRM-based retrieval, the final recommendations are also constrained by retrieved candidate items.
Such an approach fails to leverage the generative nature of LLMs to fully explore the whole item space.
{LLM as RS}~\cite{p5,howtoindex,lc-rec,eager-llm} directly utilizes LLMs for recommendation without any TRMs.
This approach typically first adapts LLMs to the recommendation task through supervised fine-tuning and then prompts them to directly generate recommendations.
However, to keep LLMs up-to-date with the evolving item pool, they need to be continuously fine-tuned, which can be prohibitively expensive.
To summarize, the three approaches do not fully harness the complementary strengths of LLMs (\eg world knowledge and reasoning) and TRMs (\eg recommendation-specific knowledge and efficiency) and are limited to \textit{shallow} exploration of the item space.

To address this, we aim to develop an LLM-based RS that can \textit{deeply} explore the item space.
Our approach is inspired by the remarkable success of LLM-based agents~\cite{agent_survey1,agent_survey2}, which can collaborate with external tools to accomplish complex tasks.
In particular, the recently proposed agent, Deep Research~\cite{deepresearch}, aims to solve complex tasks by searching the Internet.
It further promotes LLMs to leverage reasoning for autonomous interaction with tools like search engines through reinforcement learning~(RL).
In the context of recommendation, TRMs can retrieve a subset of items from the whole item space according to user interaction history, which is similar to the role of a search engine in web browsing.
Thus, TRMs can also be regarded as \textit{tools}, then LLMs can employ TRMs to retrieve relevant items from the item space for recommendation.
Motivated by this analogy, we further extend the interaction between LLMs and TRMs to \textit{multi-turn} and leverage the \textit{reasoning} capability of LLMs for deep exploration of the item space.
In contrast, existing LLM-based RSs are all limited to \textit{one-time} recommendation, which are inherently {shallow} in their exploration of the item space.

To this end, in this paper, we propose a novel LLM-based RS to perform autonomous multi-turn reasoning-retrieval between LLMs and TRMs for deep exploration of the item space, namely \textbf{\method}.
At each turn, LLMs first perform reasoning over the user interaction history and items retrieved from TRMs to generate a description about user preference.
Then, TRMs utilize the generated user preference to retrieve relevant items from the item space for LLMs.
Actually, the first step is similar to the method of TRM-enhanced LLM, while the second step is similar to that of LLM-enhanced TRM.
After autonomous multi-turn interaction, LLMs perform ranking over retrieved items to give the final recommendation results.
Compared with the method of LLM as RS, our approach avoids frequent tuning of LLMs for the update of the item space, thereby achieving higher efficiency.

Since it is difficult to label high-quality data for supervised learning, we leverage RL and propose novel designs from three aspects: \emph{recommendation model based data rollout}, \emph{recommendation-oriented hierarchical rewards}, and \emph{two-stage RL training strategy}.
For data rollout, we first introduce a preference-aware recommendation model as the TRM, which accepts textual user preferences and user interaction history as input to make recommendations.
Then, this TRM can be involved in interaction with LLMs for sequential recommendation.
The resulting interaction trajectories are collected for subsequent RL training.
In addition, we perform TRM-based data selection to control the difficulty of training data.
For reward, to optimize both the interaction process and final recommendation results, we propose a hierarchical reward function.
At the process level, we design generation format, invocation count, and preference diversity rewards for the precision, rationality, and effectiveness of the interaction between LLMs and TRMs.
At the outcome level, we design both point-wise and list-wise rewards for the performance of recommendation.
Point-wise rewards aim at the quality of each individual item.
To alleviate the sparsity, we propose to leverage textual and collaborative similarity for rewarding.
List-wise rewards aim at the quality of the overall item list.
Here, we use hit and rank metrics adapted from recommendation.
For RL training, we propose a two-stage training strategy, where the first cold-start stage aims to internalize the interaction patterns with TRMs into LLMs, while the second recommendation-oriented stage focuses on performance improvement.

To validate the effectiveness of our approach, we conduct experiments on public sequential recommendation datasets.
Experimental results show that our approach outperforms several competitive baseline models. 
Our main contributions are summarized as follows:

$\bullet$ To the best of our knowledge, it is the first time that autonomous multi-turn interaction between LLMs and TRMs has been introduced for deep exploration of the item space in recommendation.

$\bullet$ We propose novel designs for RL-based optimization from the aspects of data rollout, reward function, and training strategy.

$\bullet$ We conduct extensive experiments on public datasets, demonstrating the superiority of our approach over both traditional and LLM-based recommendation baselines.

\section{Methodology}
\label{sec:methodology}

\begin{figure*}[t]
\centering
\includegraphics[width=\linewidth]{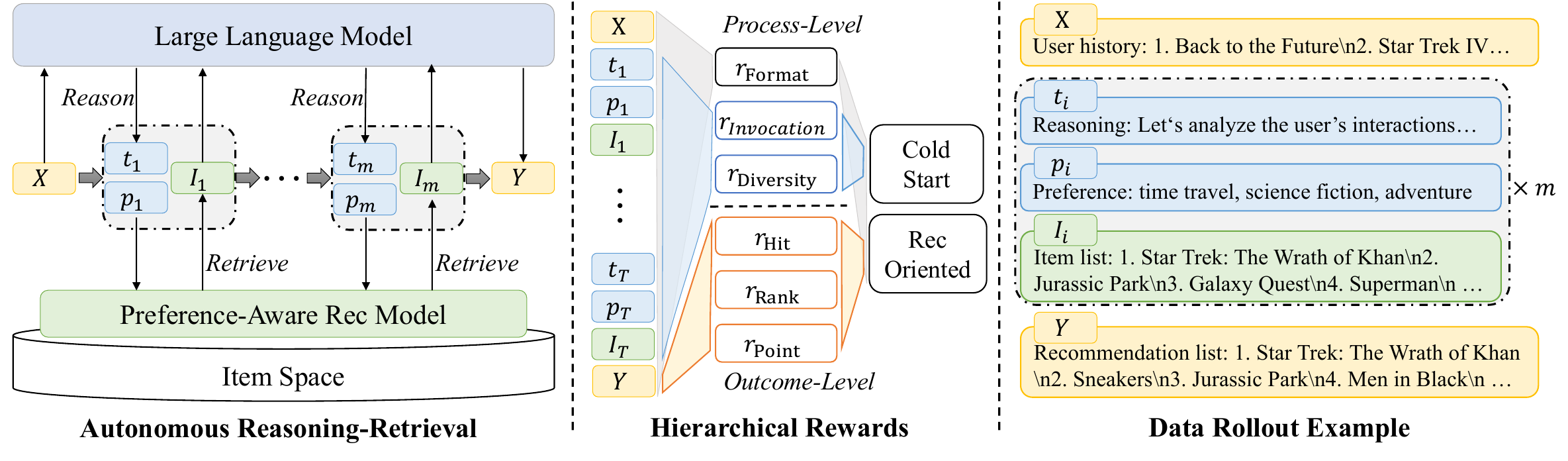}
\caption{Overview of our approach \method. 
$X$ is the user’s interaction history input to the LLM, and $Y$ is the final recommendation list output by the LLM.
$t_i$ and $p_i$ denote the thought and the user preference generated by the LLM at the $i$-th turn, respectively. $I_i$ represents the items retrieved by the TRM based on $p_i$.}
\label{fig:model}
\end{figure*}

In this section, we present our proposed LLM-based RS \textbf{\method}.
The overall architecture of our approach is illustrated in Figure~\ref{fig:model}.

\subsection{Overview of the Approach}

\paratitle{Task Formulation.}
Sequential recommendation aims to capture the user's dynamic preferences from his/her historical behaviors and predict the next potential item.
Formally, given the item set $\mathcal{V}$, a user’s historical interactions are represented as an item sequence $X = [v_1, \dots, v_n]$ in chronological order.
The recommendation task is to predict the next item $y$ based on user historical interactions.

\paratitle{LLM-based Recommendation Systems.}
Existing LLM-based recommendation systems~(RSs) for sequential recommendation can be divided into three main categories according to the relationship between LLMs and traditional recommendation models~(TRMs).

$\bullet$~\textit{LLM-enhanced TRM.}
This approach utilizes TRMs to make the recommendation and introduces LLMs as auxiliary components to help TRMs~\cite{kar,unisrec,DBLP:journals/corr/abs-2409-13545}.
Since TRMs are trained only on observed user interaction data, they do not have enough factual knowledge about each interacted item.
In contrast, LLMs are trained on massive textual corpora with abundant world knowledge.
Therefore, LLMs can provide extra features about items or augment user interaction data for TRMs.
Formally, TRMs accept user interaction history $X$ and augmented features or data $A$ from LLMs as the input to predict the next item $y$, which can be represented as follows:
\begin{equation}
    \text{Pr}(y | X) = \text{Pr}_{\text{\scriptsize TRM}}(y | X, A) \cdot \text{Pr}_{\text{\scriptsize LLM}}(A | X)
\end{equation}

$\bullet$~\textit{TRM-enhanced LLM.}
Different from LLM-enhanced TRM, this approach utilizes LLMs to make the final recommendation and introduces TRMs as auxiliary components~\cite{llamarec,llmrank}.
Since LLMs excel at complex reasoning and TRMs specialize in efficient matching, the common practice is to first use TRMs to retrieve a subset of candidate items and then use LLMs to rank these candidates for recommendation.
Formally, LLMs accept user interaction history $X$ and retrieved items $I$ from TRMs as the input to predict the next item $y$, which can be represented as follows:
\begin{equation}
    \text{Pr}(y | X) = \text{Pr}_{\text{\scriptsize LLM}}(y | X,I) \cdot \text{Pr}_{\text{\scriptsize TRM}}(I | X)
\end{equation}

$\bullet$~\textit{LLM as RS.}
Different from the above two approaches, it uses LLMs for end-to-end recommendation.
To ensure the validity of generation, this approach typically introduces concise and standardized item identifiers (\eg semantic IDs~\cite{howtoindex,lc-rec,eager-llm}) into the vocabulary of LLMs and utilizes several tokens to represent one item.
Formally, LLMs accept user interaction history $X$ as the input to directly predict the identifier of the next item $y$, which can be represented as follows:
\begin{equation}
    \text{Pr}(y | X) = \prod_{i=1}^{|y|} \text{Pr}_{ \text{\scriptsize LLM}}(y_i | X, y_{<i}),
\end{equation}
where $y_i$ and $y_{<i}$ are the $i$-th token and the tokens preceding the $i$-th position, respectively.

\paratitle{Autonomous Interaction Between LLMs and TRMs.}
Our approach is inspired by the recent LLM-based agent Deep Research~\cite{deepresearch}, where LLMs perform autonomous interaction with various tools to solve complex tasks.
In our approach, we aim to {deeply} explore the item space through \textit{autonomous interaction between LLMs and TRMs}.
As the basis, we detail this novel paradigm (Section~\ref{sec:paradigm}), where LLMs \textit{reason} to invoke TRMs and TRMs \textit{retrieve} to provide feedback for LLMs.
To develop our LLM-based RS for this paradigm, we leverage RL-based optimization and propose novel designs from three aspects: recommendation model based data rollout (Section~\ref{sec:rollout}), recommendation-oriented hierarchical reward (Section~\ref{sec:reward}), and two-stage RL training strategy (Section~\ref{sec:training}).

\subsection{Autonomous Reasoning-Retrieval For Deep Exploration of the Item Space}
\label{sec:paradigm}
\paratitle{Multi-Turn Retrieval.}
Existing LLM-based RSs are all limited to one-time recommendation, which are inherently shallow in their exploration of the item space.
To address this, our idea is to extend the interaction between LLMs and TRMs to a \textit{multi-turn retrieval} paradigm for \textit{wider} exploration.
Specifically, we regard TRMs as tools to connect LLMs with the item space.
At each turn of interaction, LLMs first generate a description about user preference according to the user interaction history and retrieved items from TRMs.
Then, TRMs perform retrieval to find items relevant to the generated user preference.
After multi-turn interaction, LLMs perform ranking over all the retrieved items to give the final recommendation results.
Formally, given the user interaction history $X$, our LLM-based RS predicts the next item $y$ as follows:
\begin{align}
\text{Pr}(y|X) = & \prod_{i=1}^{m} \underbrace{\text{Pr}_{\text{\scriptsize LLM}}(p_{i}|X,\{p_{j},I_{j}\}_{j=1}^{i-1}) \cdot \text{Pr}_{\text{\scriptsize TRM}}(I_{i}|X,p_{i})}_{\text{multi-turn retrieval}} \notag \\ 
                & \cdot \underbrace{\text{Pr}_{\text{\scriptsize LLM}}(y|X,\{p_{i}, I_{i}\}_{i=1}^m)}_{\text{rank}},
\end{align}
where $m$ is the number of interaction turns, $p_i$ and $I_i$ are the user preference generated by LLMs and the items retrieved by TRMs at the $i$-th turn, respectively.

\paratitle{Autonomous Reasoning-Retrieval.}
With the progression of interaction, the contextual information for LLMs becomes more and more complex, which requires reasoning for effective user preference generation.
Therefore, to promote \textit{deeper} exploration of the item space, we leverage the reasoning capability of LLMs and further propose the \textit{autonomous reasoning-retrieval} paradigm.
Compared with the preceding paradigm, we require LLMs to generate thoughts before user preference to conduct an analysis about the user interaction history and retrieved items in the previous turns.
Furthermore, with the reasoning capability, we can let LLMs autonomously decide when to stop the interaction, which can achieve a balance between performance and efficiency.
Formally, this paradigm can be represented as follows:
\begin{align}
    \text{Pr}(y|X) = & \prod_{i=1}^{m} \underbrace{\text{Pr}_{\text{\scriptsize LLM}}(t_i,p_i|X,\{t_j,p_j,I_j\}_{j=1}^{i-1}) \cdot \text{Pr}_{\text{\scriptsize TRM}}(I_i|X,p_i)}_{\text{autonomous reasoning-retrieval}} \notag \\ 
                             & \cdot \underbrace{\text{Pr}_{\text{\scriptsize LLM}}(y|X,\{t_i,p_i,I_i\}_{i=1}^m)}_{\text{rank}},
\end{align}
where $t_i$ is the thought generated by LLMs at the $i$-th turn.

\subsection{Recommendation Model Based Data Rollout}
\label{sec:rollout}

In the above part, we propose an autonomous reasoning-retrieval paradigm for deep exploration of the item space.
However, it is difficult to label high-quality data demonstrating the appropriate interaction between LLMs and TRMs for supervised learning.
A similar problem is also encountered by LLM-based agents, which is solved by RL~\cite{long_cot_survey}.
Inspired by this, we leverage RL to develop our LLM-based RS for the proposed paradigm.
In this part, we first introduce how to construct data for RL, including a preference-aware recommendation model, interaction data rollout, and difficulty-based data selection.

\paratitle{Preference-Aware Recommendation Model.}
Recall that in our approach, LLMs generate the user preference to invoke TRMs for item retrieval.
To support this, a straightforward method is to treat text information of items (\eg titles and descriptions) as documents and apply standard document retrieval techniques, such as BM25~\cite{bm25} or dense retrieval models~\cite{dr_survey}.
However, this method is limited by its inability to incorporate the collaborative knowledge intrinsic to recommendation, and its effectiveness is further constrained by the often suboptimal quality of item textual data. 
Therefore, we propose to integrate the user preference into a TRM (\eg SASRec~\cite{sasrec}), thereby adapting it into a preference-aware TRM.
Specifically, we first utilize a TRM to encode user interaction history $X$ and a text representation model to encode generated user preference $p_i$, then average them to derive the preference-aware user presentation $\bm{h}$, which can be represented as follows:
\begin{equation}
    \bm{h} = \frac{1}{2}\left[f_{\text{\scriptsize TRM}}(X) + f_{\text{\scriptsize Text}}(p_i)\right],
\end{equation}
where $f_{\text{\scriptsize TRM}}(X)$ and $f_{\text{\scriptsize Text}}(p_i)$ are the representations of user interaction history and generated user preference, respectively.
Then, the item prediction method remains consistent with that of the TRM, typically employing cosine similarity or the dot product with item embeddings.
Although our design is straightforward, it effectively makes the retrieval results aware of the user preference generated by LLMs.
In addition, it is model-agnostic, which can be applied to various TRMs.

\paratitle{Interaction Data Rollout.}
Based on the proposed preference-aware TRM, LLMs can interact with it, and we collect the interaction trajectories for RL.
Specifically, we design a system prompt for LLMs to follow our proposed autonomous reasoning-retrieval paradigm, which is shown as follows.
\begin{center}
\label{prompt}
\begin{tcolorbox}[
    colback=black!5!white,
    colframe=codegray,
    width=0.45\textwidth,
    title={System Prompt For LLMs}
]
{
    You are an assistant to make item recommendations for users.
    Note that you have no access to the candidate items.
    So, you must invoke a recommendation model to retrieve items you want to recommend.
    The recommendation model takes the description of user preferences as input and returns a list of relevant items as output.
    The user provides a sequence of previously interacted items, ordered by interaction time.
    To make recommendations, you need to reason about the interaction history and generate user preferences in the format of ``<|begin\_of\_preference|> $\cdots$ <|end\_of\_preference|>'' to invoke the recommendation model.
    The retrieved items will be returned in the format: ``<|begin\_of\_item\_list|> $\cdots$ <|end\_of\_item\_list|>''.
    After performing multi-turn reasoning and retrieval, you should output the final ranked list of item titles, containing the top-$K$ results.
    The reasoning process and the final recommendation list should be enclosed within ``<think> $\cdots$ </think>'' and ``<recommendation\_list> $\cdots$ </recommendation\_list>'' tags, respectively.
}
\end{tcolorbox}
\end{center}
Based on this system prompt, we input the user interaction history and provide a tag ``<think>'' at the beginning of the response to initialize the generation.
When the tag ``<|begin\_of\_preference|>'' is generated, it indicates that LLMs will generate the user preference to invoke the TRM.
Upon generating the tag ``<|end\_of\_preference|>'', we stop the generation, extract the user preference between tags, and utilize TRMs to perform retrieval with the extracted user preference.
Then, we encapsulate the retrieved items within the tags of ``<|begin\_of\_item\_list|>'' and ``<|end\_of\_item\_list|>'' and concatenate them at the end of generation.
Next, LLMs perform reasoning over the retrieved items and the initial user interaction history to decide whether to make another invocation of the TRM or the final candidate ranking.
After ranking, LLMs will generate the tag ``</think>'' to indicate the end of interaction with TRMs and the final ranked item list between the tags of ``<recommendation\_list>'' and ``</recommendation\_list>''.

\paratitle{Difficulty-Based Data Selection.}
Recent work demonstrates that the difficulty of data is important to the effectiveness of RL~\cite{kimi-k1.5}.
For our autonomous reasoning-retrieval paradigm, since LLMs perform ranking over the items retrieved by TRMs to make the final recommendation, it is important that the data have a balanced difficulty for TRMs.
To measure the difficulty, we utilize the rank of the label item predicted by our TRM.
We discard samples that are excessively difficult, that is, the rank is higher than 100.

\subsection{Recommendation-Oriented Hierarchical Rewards}
\label{sec:reward}
The reward function represents the training objective for RL.
Recently, rule-based reward~\cite{deepseek-r1, kimi-k1.5} has received extensive attention, which is easy to develop and hard to hack.
Considering these advantages, we adopt this for our approach.
To optimize both the interaction process and final recommendation results, we propose a hierarchical reward function, including process and outcome levels.
In this part, we detail the reward from these two levels.

\subsubsection{Process-Level Rewards}
\label{sec:rea_reward}
The goal of process-level reward is to provide supervision signals for the generation process of LLMs, including the interaction between LLMs and TRMs.
We design several rewards to optimize the precision, rationality, and effectiveness of interaction.

\paratitle{Generation Format Reward.}
As the basis, the generated text should adhere to the format defined in the system prompt (Section~\ref{sec:rollout}).
Specifically, we check the following points:

$\bullet$ The interaction process and final recommendation results should be enclosed within the tag pairs ({<think>}, {</think>}) and ({<recommendation\_list>}, {</recommendation\_list>}), respectively. 
Furthermore, the recommendation results should consist of a specified number of items separated by line breaks.

$\bullet$ LLMs should encapsulate user preference within the tag pair ({<|begin\_of\_preference|>}, {<|end\_of\_preference|>}). 

$\bullet$ To alleviate hallucination, LLMs must invoke TRMs at least once before making the final recommendation and cannot generate items by themselves.

According to whether the generated text meets the above requirements, we give the generation format reward, which is defined as follows:
\begin{align}
r_{\text{\scriptsize Format}} = 
\begin{cases}
    0,  & \text{if all the format points are correct} \\
    -1, & \text{otherwise}
\end{cases}
\label{eq:r_format}
\end{align}

\paratitle{Invocation Count Reward.}
In our approach, LLMs need to invoke TRMs multiple times to improve the quality of recommendation.
However, the tool invocation capability of LLMs is limited after pre-training~\cite{torl}.
Therefore, we need to encourage reasonable invocation of TRMs while preventing arbitrary invocation.  
To optimize the rationality of invoking TRMs, we design an invocation count reward, which can be represented as follows:
\begin{align}
r_{\text{\scriptsize Invocation}} = 
\begin{cases}
    1, & \text{if } m > M \\
    (m-1)*0.5, & \text{if } 1 < m \leq M \\
    0, & \text{if } m \leq 1
\end{cases}
\label{eq:r_retrieval}
\end{align}
where $m$ is the number of invocation counts and $M$ is the upper bound.

\paratitle{Preference Diversity Reward.}
Since the retrieval of TRMs is influenced by the user preference generated by LLMs, the quality of user preference is important for effective exploration of the item space.
As an extreme case, if all the generated user preferences are the same, the results of each retrieval will be the same, and exploration will fail.
To prevent this, we design a preference diversity reward to encourage the difference among preferences, thus promoting deeper exploration.
Specifically, to measure the diversity among user preferences, we encode them using a text representation model and calculate the cosine similarity between each pair.
Formally, let $m$ denote the number of user preferences in one generation, $s_{i,j}$ denote the pairwise similarity between the $i$-th and $j$-th user preferences, the reward can be represented as follows:
\begin{align}
r_{\text{\scriptsize Diversity}} &= 1 - \frac{1}{Z} \sum_{i=1}^m\sum_{j=i+1}^m s_{i,j},
\label{eq:r_div}
\end{align}
where $Z$ is the number of total preference pairs.

\subsubsection{Outcome-Level Rewards}
\label{sec:rec_reward}
The goal of outcome-level reward is to provide supervision signals for the final recommendation of LLMs.
Inspired by the evaluation metrics in recommendation, we design point-wise and list-wise rewards for comprehensiveness.

\paratitle{Point-Wise Reward.}
It aims at the quality of each individual item.
However, recommendation data are highly sparse, and only a few items have labels for rewarding.
To alleviate this, we propose a model-based point-wise reward, which leverages the textual embedding from a text representation model and collaborative embeddings from our preference-aware TRM (defined in Section~\ref{sec:rollout}).
Specifically, given the label $y$ and a predicted item $i_k$ ranked at the $k$-th position, we first utilize a text representation model to encode their textual information (\ie title) as the textual embeddings and our TRM to look up their embeddings as the collaborative embeddings.
Then, we calculate the textual and collaborative similarity score $s^t_k$ and $s^c_k$ respectively by cosine similarity.
In addition, we assign a decreasing weight $w_k$ according to the position $k$.
Finally, we calculate the weighted average of each item in the predicted item list to derive the reward, which can be represented as follows:
\begin{align}
\label{eq:r_point}
r_{\text{\scriptsize Point}} &= \frac{\sum_{k=1}^K w_k \cdot (s^t_k+s^c_k)}{2 \cdot \sum_{k=1}^K w_k}, \\
w_k &= (K - k +1)^2, \notag
\end{align}
Where $K$ is the number of predicted items.

\paratitle{List-Wise Reward.}
It aims at the quality of the overall item list.
Here, we adapt metrics from recommendation for this.

$\bullet$~\textit{Hit.}
This metric judges whether the label $y$ occurs in the generated item list $L$, which can be represented as follows:
\begin{align}
r_{\text{\scriptsize Hit}} = 
\begin{cases}
    1, & \text{if } y \in L \\
    0. & \text{otherwise} 
\end{cases}
\label{eq:r_hit}
\end{align}

$\bullet$~\textit{Rank.}
This metric further judges whether the ranking of the label $k_{y}$ is at the top positions in the generated item list $L$.
Here, we do not employ common metrics like NDCG, as it exhibits uneven positional differences, produces relatively small values, and performs poorly in our preliminary experiments.
Instead, we propose a linear decay reward, which can be represented as follows:
\begin{align}
r_{\text{\scriptsize Rank}} = 
\begin{cases}
    (K-k_y+1)*0.2, & \text{if } y \in L \\
    0, & \text{otherwise} 
\end{cases}
\label{eq:r_rank}
\end{align}
where $K$ is the number of predicted items.

\subsection{Two-Stage RL Training Strategy}
\label{sec:training}
With training data and the reward function defined in the preceding parts, we now describe how to conduct RL training.
Since we introduce a novel paradigm that LLMs are not familiar with, we add a cold-start training stage to internalize this paradigm into LLMs.
After this, we conduct normal RL training to optimize the performance.
Both stages employ the Reinforce++ algorithm~\cite{reinforce}.

\paratitle{Cold-Start RL.}
This stage utilizes process-level rewards, including generation format reward (Eq.~\ref{eq:r_format}), invocation count reward (Eq.~\ref{eq:r_retrieval}), and preference diversity reward (Eq.~\ref{eq:r_div}).
We do not add outcome-level rewards, as we observe in our early experiments that their presence does not affect the final performance but significantly interferes with the convergence and stability.
The reward function at this stage is defined as follows:
\begin{align}
r_{\text{\scriptsize Cold}} &= r_{\text{\scriptsize Format}} + r_{\text{\scriptsize Invocation}}  + r_{\text{\scriptsize Diversity}}.
\end{align}

\paratitle{Recommendation-Oriented RL.}
This stage aims to further optimize the performance of recommendation.
To achieve this, we utilize outcome-level rewards, including point-wise reward (Eq.~\ref{eq:r_point}) and list-wise reward (Eq.~\ref{eq:r_hit} and \ref{eq:r_rank}).
In addition, we add the generation format reward (Eq.~\ref{eq:r_format}) to ensure compliance with the interaction format.
The reward function at this stage is defined as follows:
\begin{align}
r_{\text{\scriptsize Rec}} &= r_{\text{\scriptsize Format}} + r_{\text{\scriptsize Point}} + r_{\text{\scriptsize Hit}} + r_{\text{\scriptsize Rank}}.
\end{align}

\section{Experiment}

In this section, we conduct empirical experiments and in-depth analyses on two public datasets to evaluate the effectiveness of our proposed \method.

\subsection{Experiment Setup}

\begin{table}[t]
    \centering
    \caption{Statistics of the preprocessed datasets.}
    \begin{tabular}{lrrrrr}
    \toprule
     Dataset    &\# Users   &\# Items   &\# Interactions & Sparsity   \\
     \midrule
     ML-1M &6,037  &3,533  &575,280  &97.303\%  \\
     Game &92,070  &25,585  &625,988  &99.972\% \\
     \bottomrule
    \end{tabular}
    \label{tab:data_statistics}
\end{table}

\subsubsection{Datasets}
To evaluate the effectiveness of our framework, we conduct experiments on two widely used public datasets for recommender systems:
(1) MovieLens-1M~\cite{ml-1m} (in short, ML-1M) contains anonymous movie ratings from users who joined MovieLens in 2000, which is a popular movie recommendation dataset.
(2) Video Games subset (in short, Game) in the latest Amazon 2023 review dataset~\cite{amazon2023} consists of user review and rating data spanning from May 1996 to September 2023, which is an e-commerce recommendation dataset.
To ensure data quality, we filter out low-activity users and items with less than five interaction records and select positive interactions with a rating greater than 3.
Afterwards, we group the historical interactions by user and sort them in chronological order, with a maximum sequence length of 20 items.
The title of the movie/product is used as descriptive text to identify each item.
Detailed statistics of the preprocessed datasets are shown in Table~\ref{tab:data_statistics}.

\subsubsection{Baseline Models}
The following competitive baseline methods are adopted for comparison with the proposed framework:

\noindent \textbf{(1) Traditional Methods}:

$\bullet$ {\textbf{BERT4Rec}}~\cite{bert4rec} utilizes a bidirectional self-attention model with a masked prediction objective for item sequence modeling.

$\bullet$ {\textbf{SASRec}}~\cite{sasrec} employs a unidirectional self-attention network to model user historical behaviors.

$\bullet$ {\textbf{FMLP-Rec}}~\cite{fmlp-rec} proposes an all-MLP architecture with learnable filters to suppress noise and capture user preferences.

$\bullet$ {\textbf{TedRec}}~\cite{tedrec} performs sequence-level text-ID semantic fusion in the frequency domain, thereby enhancing sequential recommendation performance.

\noindent \textbf{(2) LLM-based Methods}:

$\bullet$ {\textbf{LlamaRec}}~\cite{llamarec} introduces a two-stage framework that first performs ID-based retrieval and then employs a fine-tuned LLM for re-ranking to enhance performance.

$\bullet$ {\textbf{LC-Rec}}~\cite{lc-rec} designs various alignment tuning tasks to enhance the integration of collaborative semantics in LLMs.

$\bullet$ {\textbf{LLMs$_{\text{+Prompt}}$}} employs prompt engineering to enable LLMs to emulate the paradigm of \method. Through multi-turn interaction with TRMs, LLMs iteratively explore the item space and subsequently generate the final recommendation.

$\bullet$ {\textbf{LLMs$_{\text{+TRM}}$}} adopts TRM-enhanced LLM approach. First, a traditional recommender is employed to acquire a small set of candidate items, and then LLMs are instructed to re-rank retrieved candidates.

For LLMs$_{\text{+Prompt}}$ and LLMs$_{\text{+TRM}}$ methods, we tried various LLMs as backbones, including Qwen-2.5-14B-Instruct, Qwen-2.5-32B-Instruct~\cite{qwen2.5}, and DeepSeek-R1-Distill-Qwen-32B~\cite{deepseek-r1}.

\subsubsection{Evaluation Settings}

To evaluate the model performance in sequential recommendation, we leverage two widely used metrics: top-$K$ Recall and Normalized Discounted Cumulative Gain (NDCG), with $K$ set to 5 and 10.  
Following prior studies~\cite{sasrec,s3rec}, we employ the \emph{leave-one-out} strategy to split the dataset into training, validation, and test sets.
Specifically, for each user interaction sequence, the most recent item is used as the test sample, the second most recent item is used for validation, and the remaining items are used for training.
To ensure a rigorous comparison, we treat all items that a user has not interacted with as potential candidates and perform evaluation over the entire item space to eliminate sampling bias.

\subsubsection{Implementation Details}

We adopt Qwen2.5-7B-Base~\cite{qwen2.5} as the backbone LLM for our proposed \method approach. 
For RL training, we adopt the Reinforce++ algorithm~\cite{reinforce} and develop our method based on the OpenRLHF framework~\cite{openrlhf}.
Regarding hyperparameters, we use the AdamW optimizer with a learning rate of 1e-6. 
Additionally, we set the KL coefficient to 0 and the temperature to 1 to enhance model exploration. 
The mini-batch size is set to 512, the rollout batch size is set to 128 and 8 examples are generated for each prompt.
The upper bound $M$ of the invocation count reward is set to 3.
We train the model for 20 steps in the cold-start stage and 80 steps in the recommendation-oriented stage.
For the preference-aware TRM, we adopt TedRec~\cite{tedrec} with an embedding size of 128 as the base TRM.
For each time of retrieval, we return 20 relevant items to LLMs.
The same model is also used for LLMs$_{\text{+Prompt}}$ and LLMs$_{\text{+TRM}}$ baseline methods to ensure fair comparison.
We leverage BGE-large~\cite{bge} as the text representation model to encode the preference text or item title for similarity calculation in the reward functions (\ie Eq.~\ref{eq:r_div} and ~\ref{eq:r_point}).
We implement all traditional baseline methods based on RecBole~\cite{recbole}, and refer to the original paper for hyperparameter grid search to obtain the best performance.
For LC-Rec, we set the beam size to 50 during the inference process.

\begin{table*}[t]
\centering
\caption{Performance comparison of different recommendation methods. The best and second-best performances are indicated in bold and underlined font, respectively. }
\label{tab:main_res}
\begin{tabular}{lcccccccc}
\toprule
\multirow{2}{*}{Methods} & \multicolumn{4}{c}{ML-1M}               & \multicolumn{4}{c}{Game}                \\
\cmidrule(l){2-5} \cmidrule(l){6-9}
                         & Recall@5 & Recall@10 & NDCG@5 & NDCG@10 & Recall@5 & Recall@10 & NDCG@5 & NDCG@10 \\
\midrule
\midrule
\multicolumn{9}{c}{\textit{Traditional Methods}}     \\
\midrule
\midrule
BERT4Rec                                     & 0.1485	& 0.2304	& 0.0986	& 0.1250	& 0.0420	& 0.0667	& 0.0270	& 0.0349   \\
SASRec                                       & 0.1501 	& 0.2309 	& 0.0100 	& 0.1252 	& 0.0579	& 0.0898	& 0.0328	& 0.0431  \\
FMLP-Rec                                     & 0.1524	& \underline{0.2314}	& 0.0999	& 0.1244	& 0.0519	& 0.0831	& 0.0333	& 0.0433 \\
TedRec                                       & \underline{0.1550}   & 0.2312    & \underline{0.1019} & \underline{0.1273}  & 0.0598   & \underline{0.0929}    & 0.0387 & 0.0501  \\
\midrule
\midrule
\multicolumn{9}{c}{\textit{LLM-based Methods}}          \\
\midrule
\midrule
LlamaRec                                     & 0.1353   & 0.2032    & 0.0846 & 0.1067  & 0.0591   & 0.0925    & 0.0369 & 0.0476  \\
LC-Rec                                       & 0.1501   & 0.2237    & 0.0993 & 0.1210   & \underline{0.0601}   & 0.0914    & \underline{0.0400}   & 0.0501  \\
\midrule
Qwen-2.5-14B$_{\text{+Prompt}}$      &   0.0825	& 0.1396	& 0.0523	& 0.0706    & 0.0393   & 0.0662    & 0.0257 & 0.0344  \\
Qwen-2.5-32B$_{\text{+Prompt}}$    &       0.1081	& 0.1741	& 0.0683	& 0.0895  & 0.0443   & 0.0764    & 0.0279 & 0.0382  \\
R1-Distill-Qwen-32B$_{\text{+Prompt}}$  &  0.0165	& 0.0267	& 0.0099	& 0.0133    & 0.0139   & 0.0231    & 0.0089 & 0.0118  \\
\midrule
Qwen-2.5-14B$_{\text{+TRM}}$                        & 0.1348   & 0.2165    & 0.0882 & 0.1144  & 0.0572   & 0.0899    & 0.0379 & 0.0496  \\
Qwen-2.5-32B$_{\text{+TRM}}$                        & 0.1426   & 0.2197    & 0.0933 & 0.1179  & 0.0597   & 0.0910     & 0.0397 & 0.0502  \\
R1-Distill-Qwen-32B$_{\text{+TRM}}$                 & 0.1179   & 0.1950     & 0.0719 & 0.0970   & 0.0600   & 0.0916    & 0.0391 & \underline{0.0505} \\
\midrule
\method                & \textbf{0.1629}	& \textbf{0.2423}	& \textbf{0.1088}	& \textbf{0.1337}	& \textbf{0.0678} 	& \textbf{0.1016} 	& \textbf{0.0453} 	& \textbf{0.0562}  \\
\bottomrule
\end{tabular}
\vspace{-5pt}
\end{table*}

\subsection{Overall Results}

We compare \method with traditional and LLM-based methods on two public recommendation datasets. The overall results are shown in Table~\ref{tab:main_res}. From these results, we can find that:

For traditional recommendation models, FMLP-Rec outperforms both SASRec and BERT4Rec by introducing a more advanced full-MLP architecture.
TedRec, which proposes sequence-level text-ID semantic fusion, demonstrates superior performance to baseline models that only involve item ID and collaborative signals.
This observation indicates that incorporating textual features as auxiliary information can significantly enhance recommendation performance.
Regarding LLM-based recommendation models, the fine-tuned LlamaRec and LC-Rec attain better results than most traditional methods on the Game dataset.
However, LlamaRec underperforms compared to LC-Rec, potentially due to the limited candidate set constraining the capabilities of LLMs.
Moreover, developing the proposed autonomous interaction paradigm via prompt engineering (\ie Qwen-2.5-14B$_{\text{+Prompt}}$, Qwen-2.5-32B$_{\text{+Prompt}}$, R1-Distill-Qwen-32B$_{\text{+Prompt}}$) often yields suboptimal results.
The reason could be that effective interaction with TRMs requires LLMs to possess stronger instruction-following and reasoning abilities.
We further observe that R1-Distill-Qwen-32B strictly replicates the reasoning patterns of DeepSeek R1, yet demonstrates limited capability in adhering to user-specified requirements.
Most TRM-enhanced LLMs (\ie Qwen-2.5-14B$_{\text{+TRM}}$, Qwen-2.5-32B$_{\text{+TRM}}$, R1-Distill-Qwen-32B$_{\text{+TRM}}$) do not exhibit significant improvements over the base recommendation model (\ie TedRec), which may be attributed to the lack of collaborative knowledge in LLMs.
An exception is R1-Distill-Qwen-32B, which surpasses TedRec in the Game dataset, likely benefiting from its advanced reasoning capabilities.

Finally, our proposed \method maintains the best performance in all cases, with substantial improvements over both traditional and LLM-based baselines.
Different from prior LLM-based recommendation methods, we propose a novel deep recommendation framework that enables LLMs to automatically interact with TRMs during reasoning for mutual enhancement. Through carefully designed recommendation-oriented hierarchical rewards and two-stage RL training, we significantly enhance the reasoning capabilities of LLMs in recommendation scenarios, resulting in more comprehensive and deliberate recommendations.

\begin{table*}[t]
\centering
\caption{Ablation study of our method.}
\label{tab:ablation}
\begin{tabular}{lcccccccc}
\toprule 
\multicolumn{1}{l}{\multirow{2}{*}{Methods}} & \multicolumn{4}{c}{ML-1M}          & \multicolumn{4}{c}{Game}                \\
\cmidrule(l){2-5} \cmidrule(l){6-9}
\multicolumn{1}{c}{}                         & Recall@5 & Recall@10 & NDCG@5 & NDCG@10 & Recall@5 & Recall@10 & NDCG@5 & NDCG@10 \\
\midrule
(0) \method                 & \textbf{0.1629}	& \textbf{0.2423}	& \textbf{0.1088}	& \textbf{0.1337}	& \textbf{0.0678} 	& \textbf{0.1016} 	& \textbf{0.0453} 	& \textbf{0.0562}  \\
\midrule
(1) \  \emph{w/o RL}                    & 0.0452	& 0.0729	& 0.0293	& 0.0382	& 0.0194	& 0.0328	& 0.0122	& 0.0165  \\
(2) \  \emph{w/o Cold-Start}                & 0.1600	& 0.2386	& 0.1068	& 0.1319	& 0.0653 	& 0.0989 	& 0.0435 	& 0.0545   \\
(3) \  \emph{w/o Two Stage RL}                & 0.1604	& 0.2397	& 0.1073	& 0.1324	& 0.0663 	& 0.1001 	& 0.0444 	& 0.0553   \\
(4) \  \emph{w/o Data Selection}                & 0.1601	& 0.2391	& 0.1070	& 0.1321	& 0.0658 	& 0.0998 	& 0.0440 	& 0.0549     \\
\bottomrule
\end{tabular}
\vspace{-10pt}
\end{table*}

\subsection{Ablation Study}

To investigate the contribution of the various components integrated in our proposed \method, we conducted ablation studies on two datasets, with the results presented in Table~\ref{tab:ablation}.
Specifically, we compared \method with the following four variants:

(1) \underline{\emph{w/o RL}} removes RL training, which is used to further enhance the autonomous interaction between LLMs and TRMs to improve the recommendation performance.
We can see that this variant is significantly inferior to \method, indicating that an LLM without specialized adaptation often struggles with the complex tasks involved in our scenarios.

(2) \underline{\emph{w/o Cold-Start}} removes cold start stage in RL training. 
The results show that this variant consistently underperforms compared to \method, which may be attributed to the absence of early-stage guidance that enables LLMs to engage in multi-turn interactions with TRMs.

(3) \underline{\emph{w/o Two Stage RL}} directly applies all the proposed rewards for single-stage RL. 
From the results, it can be observed that this naive approach leads to a decline in performance.
We speculate that the invocation count and preference diversity rewards induce reward hacking during the RL process, causing the LLM to focus on meaningless multiple retrieval or superficial preference diversification while ignoring the optimization of recommendation quality.

(4) \underline{\emph{w/o Data Selection}} removes TRM-based data selection and retains the entire training set. 
This variant has a detrimental impact on performance, possibly due to the inclusion of numerous low-quality samples where TRM predicted results outside the top 100, resulting in many rollouts getting zero reward.


\subsection{Further Analysis}

\begin{figure}[]
\centering
\includegraphics[width=0.99\linewidth]{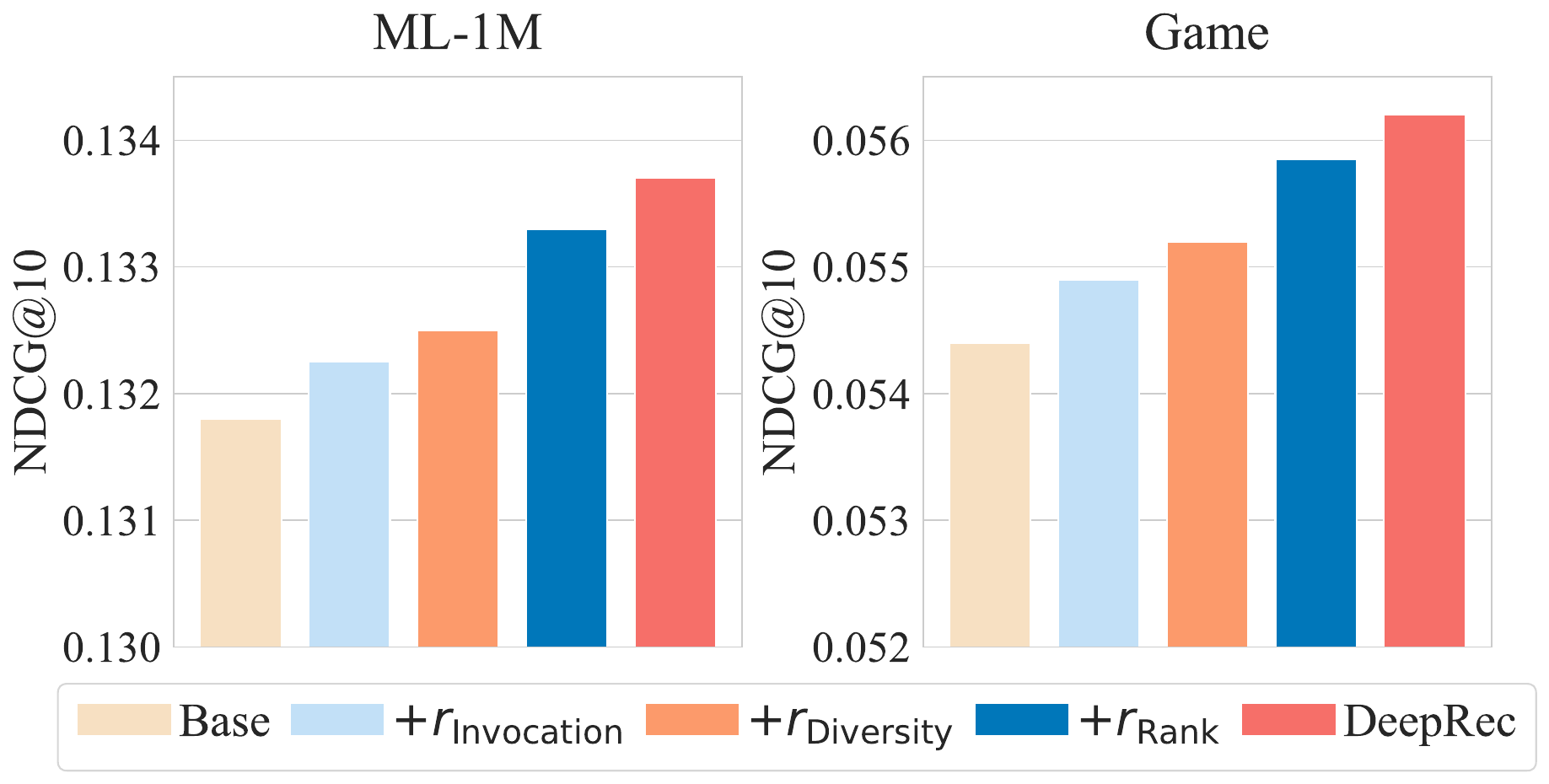}
\caption{The performance impact of different rewards.}
\label{fig:reward_ablation}
\vspace{-10pt}
\end{figure}

\subsubsection{Effects Analysis \wrt Various Rewards}
Our proposed RL algorithm consists of various recommendation-oriented rewards including:
Generation format reward \underline{$r_{\text{\scriptsize Format}}$~(Eq.~\ref{eq:r_format})} is designed to standardize the response format, thereby facilitating interaction with TRMs and effective result parsing.
Invocation count reward \underline{$r_{\text{\scriptsize Invocation}}$ (Eq.~\ref{eq:r_retrieval})} is designed to encourage LLMs to automatically interact with TRMs during the cold-start stage.
Preference diversity reward \underline{$r_{\text{\scriptsize Diversity}}$ (Eq.~\ref{eq:r_div})} is designed to guide LLMs to iteratively refine or further explore the reasoning of user personalized preferences.
Point-wise reward \underline{$r_{\text{\scriptsize Point}}$ (Eq.~\ref{eq:r_point})} measures the similarity between the predicted item and the ground-truth item, which is a continuous reward function that can effectively alleviate reward sparsity.
Hit reward \underline{$r_{\text{\scriptsize Hit}}$ (Eq.~\ref{eq:r_hit})} is designed to evaluate whether the recommendation list generated by LLMs contains the ground-truth item.
Rank reward \underline{$r_{\text{\scriptsize Rank}}$ (Eq.~\ref{eq:r_rank})} is designed to train LLM to rank the item list according to the likelihood that the user will interact with each item.
To evaluate the effectiveness of the aforementioned rewards, we conduct an in-depth analysis of their impact. 
Specifically, we begin with a base method that only employs format reward and hit reward, which ensures that LLMs can correctly interact with TRMs and complete the recommendation task.
Subsequently, we incrementally incorporate the proposed rewards into the base method, one at a time, to investigate their individual effects.
From the results presented in Figure~\ref{fig:reward_ablation}, we can see that progressively incorporating the proposed recommendation-oriented rewards into RL training leads to substantial performance improvements.

\begin{figure}[t]
\centering
\includegraphics[width=0.99\linewidth]{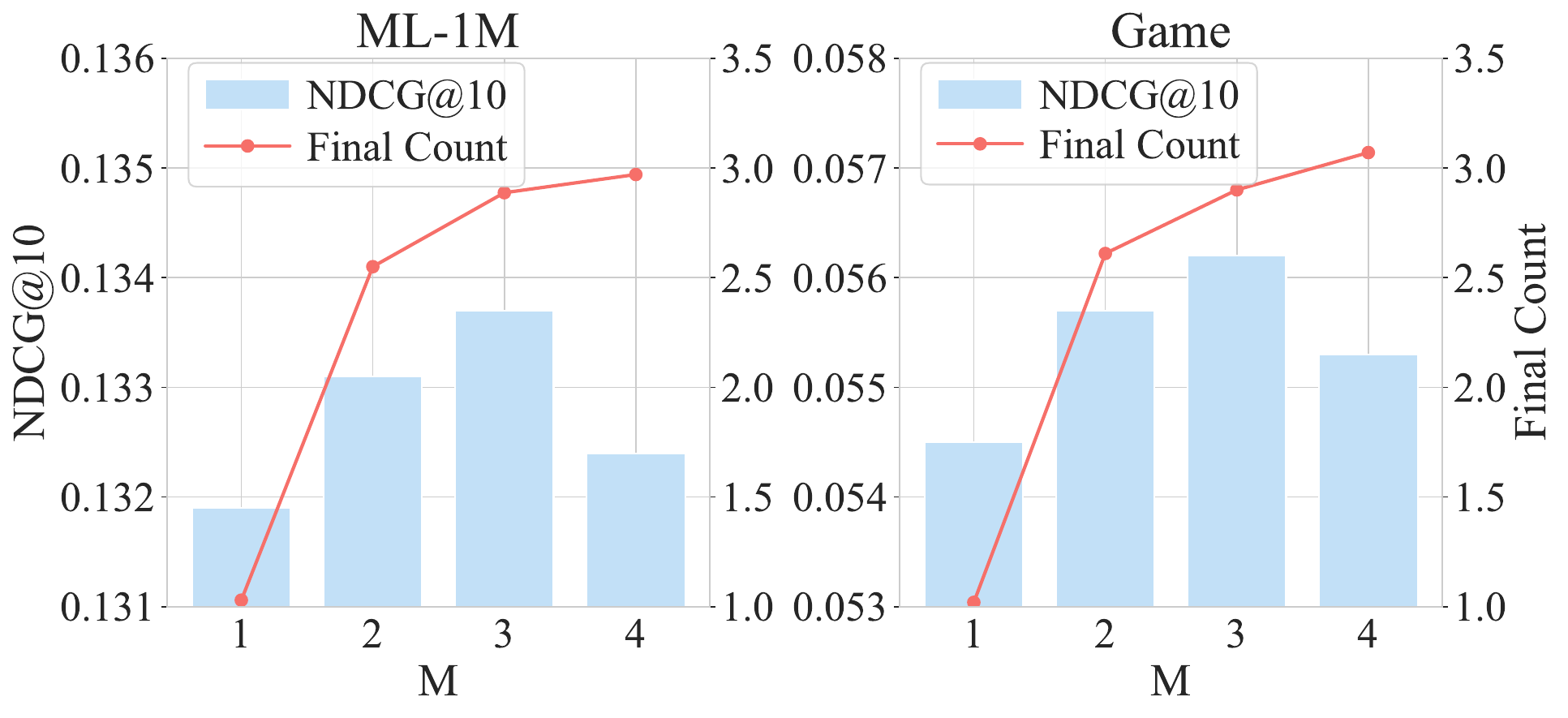}
\caption{Performance \wrt initial invocation count. 
}
\label{fig:invocation_num}
\vspace{-10pt}
\end{figure}

\subsubsection{Initial Invocation Count Analysis}

In our approach, LLMs autonomously invoke TRMs during the reasoning process to achieve deep exploration of the item space. 
Therefore, we further investigate how the initial invocation count obtained after the cold-start stage affects the recommendation performance. 
Specifically, we tune the upper limit $M$ of the invocation count reward within \{1, 2, 3, 4\}, which leads the model to perform different numbers of interactions with TRMs after cold-start RL.
As shown in Figure~\ref{fig:invocation_num}, without the invocation count reward (\ie the initial invocation count is 1), the model struggles to conduct multiple retrievals for comprehensive item space exploration.
Consequently, the final invocation count converges to 1, which hinders iterative refinement and results in significant performance degradation.
The optimal initial invocation count is 3 on both ML-1M and Game datasets.
Moreover, even when the initial number is 4, the final number of invocations tends to converge to about 3. 
However, this leads to performance degradation and increased computational cost for training.
We hypothesize that this is due to the introduction of redundant or distracting items when too many retrievals are performed.

\subsubsection{Generalize to Unseen Domain}

As described in the literature on large reasoning models, LLMs trained in mathematical or coding scenarios also demonstrate strong performance across various open-world tasks~\cite{long_cot_survey}.
Furthermore, an advantage of our proposed framework is that it does not require expensive LLM adaptation for updates in the item space.
Thus, in this section, we try to evaluate the generalization ability of \method on an unseen domain. 
Specifically, we apply the model trained on the Game dataset to another subset of the Amazon review dataset (\ie Musical Instruments). 
The results are presented in Table~\ref{tab:generalization}.
Despite not being trained on these specific domains, \method still exhibits impressive performance compared to baseline models.
This finding indicates that our approach can effectively enhance the comprehensive reasoning capabilities of LLMs in recommendation scenarios.

\begin{table}[t]
\centering
\caption{Generalization performance on unseen domain.}
\label{tab:generalization}
\resizebox{\linewidth}{!}{
\begin{tabular}{lcccc}
\toprule 
Methods& Recall@5 & Recall@10 & NDCG@5 & NDCG@10\\
\midrule
TedRec                       & \underline{0.0316}   & \underline{0.0505}   & 0.0204 & 0.0262  \\
Qwen-2.5-32B$_{\text{+TRM}}$ & 0.0305   & 0.0499    & \underline{0.0205} & \underline{0.0265}  \\
\method                      & \textbf{0.0326}   & \textbf{0.0516}    & \textbf{0.0210}  & \textbf{0.0271} \\
\bottomrule
\end{tabular}
}
\end{table}

\section{Related Work}

\paratitle{Sequential Recommendation.}
Sequential recommendation aims to model the user's dynamic behavior sequences and capture personalized preferences to predict the next potential item.
Early works~\cite{fpmc,fossil} adopt the Markov chain assumption, learning transition probabilities between items through a transition matrix.
With the rapid development of deep learning, various deep neural networks have been employed to capture sequential patterns in item ID sequences, including convolutional neural networks (CNNs)~\cite{caser}, recurrent neural networks (RNNs)~\cite{gru4rec,DBLP:conf/recsys/TanXL16}, graph neural networks (GNNs)~\cite{DBLP:conf/aaai/WuT0WXT19,DBLP:conf/ijcai/XuZLSXZFZ19}, multilayer perceptrons (MLPs)~\cite{fmlp-rec}, and Transformer-based models~\cite{sasrec,bert4rec}.
Furthermore, several studies~\cite{cl4srec,duorec} have explored the use of contrastive learning to improve sequence representation learning.
However, these approaches are primarily grounded in item IDs and often neglect the rich semantic information contained in item content (\ie, title, description, and category).
Thus, some studies have introduced item attributes~\cite{fdsa,s3rec} or pre-encoded textual features from LLMs~\cite{unisrec,vqrec} to facilitate user preference modeling. 
More recently, the generative recommendation paradigm has emerged~\cite{tiger,lc-rec}, where each item is represented as discrete tokens and target items are generated in an autoregressive manner.

\paratitle{LLM-Based Recommendation.}
Large language models (LLMs) have been widely adopted to enhance recommender systems, owing to their remarkable capabilities in semantic understanding and knowledge reasoning~\cite{llm_survey,llmrec_survey1,tallrec}.
Existing LLM-based recommendation methods can be categorized into three approaches: LLM-enhanced TRM, TRM-enhanced LLM, and LLM as RS.
LLM-enhanced TRM adopts LLMs as an auxiliary component to enhance traditional recommendation models (TRMs) through techniques such as text information encoding~\cite{unisrec,vqrec,lsvcr}, user/item feature generation~\cite{kar,llmesr}, and interaction data augmentation~\cite{DBLP:journals/corr/abs-2409-13545,DBLP:journals/corr/abs-2401-13870}.
TRM-enhanced LLM utilizes TRMs to retrieve candidate items, which are subsequently re-ranked by LLMs~\cite{llamarec,llmrank}. 
LLM as RS typically adopts an end-to-end paradigm, employing LLMs to directly complete recommendation tasks. 
To this end, researchers often represent each item using discrete tokens and apply supervised fine-tuning (SFT) to align the LLM with RS~\cite{p5,instructrec,lc-rec,e4srec,eager-llm}.
Furthermore, recent studies have explored various methods to better incorporate collaborative information into LLMs, including applying direct preference optimization~\cite{s-dpo} and refining item identifiers via collaborative signals~\cite{letter}.
Despite the progress, these methods require costly fine-tuning of LLMs to adapt to the updates in the item pool. 
In addition, they usually rely on naive and single-step reasoning to generate target items.
In contrast, our proposed \method improves recommendation performance by enabling more comprehensive and deeper exploration of the item space through multi-turn interaction between LLMs and TRMs.

\paratitle{LLM Reasoning.}
Recently, advanced large language models (LLMs), such as OpenAI o1~\cite{openai-o1} and DeepSeek R1~\cite{deepseek-r1}, have shown impressive performance on complex reasoning tasks, primarily due to their ability to reason with long Chain-of-Thought (CoT).
Different from earlier methods based on CoT prompting~\cite{long_cot_survey,cot_survey}, recent works typically incentivize the reasoning capabilities of LLMs through reinforcement learning~\cite{deepseek-r1,kimi-k1.5,qwq32b}.
For instance, DeepSeek R1 utilizes GRPO~\cite{grpo}, a value-free reinforcement learning algorithm, to facilitate self-reflection and iterative refinement before providing a response, thereby enabling DeepSeek R1 to generate more comprehensive and well-justified solutions.
To further enhance the reasoning ability of LLMs, researchers have explored various directions, including developing novel reinforcement learning algorithms~\cite{drgrpo,dapo}, as well as integrating external tools~\cite{still3,r1-searcher,torl}.
Motivated by these advancements, several studies have introduced reasoning techniques into recommender systems.
As an example, ReaRec~\cite{rearec} adopts a latent reasoning technique to enhance traditional recommendation models for next-item prediction.
Furthermore, Rec-R1~\cite{rec-r1} proposes to improve the query rewriting ability of LLM through reinforcement learning. 
However, this method employs to enhance a weak item retriever, which causes it to lag behind dedicated sequential recommendation models in performance.
In contrast, our approach introduces a TRM as the tool, and LLM can automatically interact with the TRM during reasoning to deeply explore the entire item space.

\section{Conclusion}

In this paper, we proposed \method, a novel recommendation framework that allows LLMs to autonomously interact with TRMs for deep exploration of the item space.
Unlike previous LLM-based approaches, which are often limited to one-time recommendatio and shallow item exploration, \method achieves deeper and more deliberate recommendations by incorporating a preference-aware TRM and enabling autonomous multi-turn reasoning-retrieval.
To achieve effective RL training, we designed recommendation-oriented hierarchical rewards that optimize both the interaction process and the overall performance.
Furthermore, we introduced a two-stage training strategy consisting of a cold-start stage to internalize our proposed paradigm, followed by a recommendation-oriented stage that focuses on enhancing recommendation performance.
Extensive experiments on public datasets showed the effectiveness and robustness of \method, demonstrating its superiority over traditional and LLM-based recommendation baselines. 
This work highlights the potential of integrating reasoning-oriented LLM agents with efficient retrieval tools, paving the way for more intelligent and adaptive recommender systems.
For future work, we plan to integrate multiple types of TRMs (\eg CTR models) into the automatic interaction process to further enhance LLM reasoning.


\bibliographystyle{ACM-Reference-Format}

\balance
\bibliography{ref}

\end{document}